\documentclass[aps,prl,superscriptaddress,twocolumn,showpacs,amsmath,amssymb,letter]{revtex4}

\usepackage{graphicx}
\usepackage{dcolumn}
\usepackage{bm}

\begin{document}

\title{Doping dependence of the coupling of electrons to bosonic modes
in the single-layer high-temperature Bi$_{2}$Sr$_{2}$CuO$_{6}$
superconductor}
\author{W. Meevasana}
\email[]{non@stanford.edu}
\affiliation {Department of Physics,
Applied Physics, and Stanford Synchrotron Radiation Laboratory,
Stanford University, Stanford, CA 94305}

\author{N.J.C. Ingle}
\altaffiliation{Present address: Department of Physics and
Astronomy, University of British Columbia, Vancouver, Canada}
\affiliation {Department of Physics, Applied Physics, and Stanford
Synchrotron Radiation Laboratory, Stanford University, Stanford,
CA 94305}

\author{D.H. Lu}
\affiliation {Department of Physics, Applied Physics, and Stanford
Synchrotron Radiation Laboratory, Stanford University, Stanford,
CA 94305}

\author{J.R. Shi}
\affiliation{Department of Physics, University of Texas, Austin,
TX}

\author{F. Baumberger}
\affiliation {Department of Physics, Applied Physics, and Stanford
Synchrotron Radiation Laboratory, Stanford University, Stanford,
CA 94305}

\author{K.M. Shen}
\altaffiliation{Present address: Department of Physics and
Astronomy, University of British Columbia, Vancouver, Canada}
\affiliation {Department of Physics, Applied Physics, and Stanford
Synchrotron Radiation Laboratory, Stanford University, Stanford,
CA 94305}

\author{W.S. Lee}
\affiliation {Department of Physics, Applied Physics, and Stanford
Synchrotron Radiation Laboratory, Stanford University, Stanford,
CA 94305}

\author{T. Cuk}
\affiliation {Department of Physics, Applied Physics, and Stanford
Synchrotron Radiation Laboratory, Stanford University, Stanford,
CA 94305}

\author{H. Eisaki}
\affiliation{Nanoelectronic Research Institute, AIST, Tsukuba
305-0032, Japan}

\author{T.P. Devereaux}
\affiliation {Department of Physics, University of Waterloo,
Waterloo, Ontario, Canada N2L 3G1}

\author{N. Nagaosa}
\affiliation {CREST, Department of Applied Physics, University of
Tokyo, Bunkyo-ku, Tokyo 113, Japan}

\author{J. Zaanen}
\altaffiliation{On leave of absence from the Instituut-Lorentz for
Therorectical Physics, Leiden University, Leiden, The Netherlands}
\affiliation {Department of Physics, Applied Physics, and Stanford
Synchrotron Radiation Laboratory, Stanford University, Stanford,
CA 94305}
\author{Z.-X. Shen}
\affiliation {Department of Physics, Applied Physics, and Stanford
Synchrotron Radiation Laboratory, Stanford University, Stanford,
CA 94305}

\date{\today}

\begin{abstract}
A recent highlight in the study of high-$\rm T_c$ superconductors
is the observation of band renormalization / self-energy effects
on the quasiparticles.  This is seen in the form of kinks in the
quasiparticle dispersions as measured by photoemission and
interpreted as signatures of collective bosonic modes coupling to
the electrons. Here we compare for the first time the
self-energies in an optimally doped and strongly overdoped,
non-superconducting single-layer Bi-cuprate
(Bi$_{2}$Sr$_{2}$CuO$_{6}$). Besides the appearance of a strong
overall weakening, we also find that weight of the self-energy in
the overdoped system shifts to higher energies. We present
evidence that this is related to a change in the coupling to
c-axis phonons due to the rapid change of the c-axis screening in
this doping range.
\end{abstract}

\pacs{71.38.-k, 74.72.Hs, 79.60.-i}
\maketitle

The coupling of electrons to bosonic modes is at the heart of the
mechanism of Cooper pair formation in conventional
superconductors. For high temperature superconductors, evidence of
electron-boson coupling has manifested itself in the form of a
dispersion anomaly ("kink") obtained from angle-resolved
photoemission (ARPES) experiments, leading to lively discussions
on the nature of the bosons
\cite{Group:Review,Mode:Pasha,Mode:Kaminski,Mode:Johnson,Mode:Lanzara,mode:Fink,Mode:Gromko,Norman:SpinAndTightbinding,B1gMode:Tanja,Mode:Gweon,
Bi2223:Weisheng}. These discussions focus on bosons with a sharp
energy scale, such as phonons and the magnetic resonance mode,
rather than on the continua of excitations associated with the
non-Fermi liquid nature of the normal state. An important issue is
if a single, unique boson is involved or some spectrum of bosons.
In a recent study in La$_{2-x}$Sr$_x$CuO$_4$ (LSCO), evidence was
presented in favor of the latter \cite{MultiMode:XJZhou}.

As the physical properties of cuprates change rapidly with doping,
a natural question is whether the coupling of the electrons to
bosonic modes also changes with doping.  To date, a few doping
dependent results reported from ARPES are the weakening of the
dispersion kink with doping
\cite{Mode:Johnson,Mode:Lanzara,MultiMode:XJZhou}, and the polaron
formation found in the approach to zero doping
\cite{Polaron:Kyle}.

Bi$_{2}$Sr$_{2}$CuO$_{6}$ (Bi2201) provides a good opportunity to
study this issue because (a) detailed measurements of the normal
state can be performed at low temperatures because of its low $\rm
T_c$, avoiding complications associated with the superconducting
gap, (b) its stable surface makes possible very high statistics,
which are essential for these experiments, and (c) complicating
bi-layer splitting effects are absent in this single-layer
cuprate. We report high-resolution photoemission data from
optimally-doped and strongly-overdoped, non-superconducting
Bi2201. These data reveal that the self-energy changes drastically
in this doping range. Besides an apparent reduction of the overall
strength of the self-energy by a factor of two, we find that the
self-energy changes {\em qualitatively}: in the strongly-overdoped
sample, the self-energy is clearly peaked near 75 meV, suggesting
the dominance of a mode at this energy. For the optimally doped
sample, the self-energy has significantly more weight at lower
energies, suggesting couplings to lower energy modes. This is hard
to reconcile with interpretations only involving propagating
magnetic excitations. The only available candidate for the
magnetic excitations is the magnetic resonance and it is known
that this excitation softens in the overdoped regime
\cite{MagneticResonance:Keimer}.

The measured self-energy suggests a behavior of electrons coupled
to collective modes at several specific frequencies, such as
phonons. We suggest that the large change in the self-energy in
this doping range reflects a large scale change in the
electrodynamic nature of the cuprates in this doping range. At
optimal doping, in the normal state, the cuprates show no plasmon
peak in the direction perpendicular to the planes, and therefore
are regarded as polar insulators along that direction. This
absence of a c-axis plasmon peak, which is believed to be due to
large damping effects \cite{PlasmonInOut:Marel}, implies that the
coupling between electrons and the c-axis phonons is unscreened
and therefore unusually strong \cite{smallq:Tom}. In the strongly
overdoped regime however, the c-axis conductivity becomes metallic
and therefore this coupling is expected to diminish due to the
c-axis metallic screening.  It appears that this decoupling from
the c-axis phonons might be responsible for much of the change
occurring in the self-energy in this doping range accessed in this
experiment. We show that the experimentally determined c-axis
electron energy loss function is a good model for the Eliashberg
function $\alpha^2 F(\omega)$ of the optimally doped system, and
that it continues to be a good model for the overdoped sample if
we assume that the enhanced screening due to the c-axis
metallicity involves a characteristic screening frequency,
$\omega_{src,c}\sim 60$ meV,  such that phonon modes below
$\omega_{src,c}$ can be regarded as screened.

 We have measured two sets of  single crystals of Pb-substituted
Bi2201. The optimally doped (OP) samples,
Pb$_{0.55}$Bi$_{1.5}$Sr$_{1.6}$La$_{0.4}$CuO$_{6+\delta}$, have a
$\rm T_c$ = 35 K. The overdoped (OD) samples,
Pb$_{0.38}$Bi$_{1.74}$Sr$_{1.88}$CuO$_{6+\delta}$, are
non-superconducting ($\rm T_c$ $<$ 4 K). Note that hole doping is
adjusted by changing the La and O content while Pb doping does not
change $\rm T_c$ but weakens effects of the super-lattice
structure. ARPES data were collected on a Scienta-200 analyzer at
the Stanford Synchrotron Radiation Laboratory (SSRL) Beamline 5-4
with a photon energy of 23.7 eV and a base pressure of $2 \times
10^{-11}$ torr. Additional data were also collected on a
Scienta-2002 analyzer with He I light (21.2 eV) from a
monochromated and modified Gammadata He lamp (HeLM); the pressure
was $6 \times 10^{-11}$ torr. Samples were cleaved \emph{in situ}
at the measurement temperature. The energy resolution was set to
13 meV for SSRL and 8 meV for HeLM. The average momentum
resolution at these photon energies was ~ 0.013 {\AA}$^{-1}$ (or
0.35$^{\circ}$).

\begin{figure} [t]
\includegraphics [width=3.0in, clip]{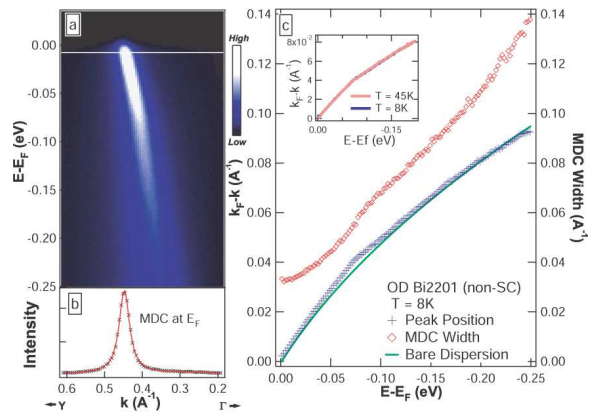}
\caption{\label{fig1:SC} (a) ARPES spectrum of OD sample,
non-superconducting (non-SC) along (0, 0) to ($\pi$, $\pi$) at T =
8 K. (b) is the fit of MDC at $E_F$. (c) Peak position, MDC width
and approximated bare dispersion for extracting Re($\Sigma$);
inset is the comparison of peak position taken at 8 K and 45 K.}
\end{figure}

The data are taken in the normal state along the nodal cut ( 0, 0)
to ( $\pi$, $\pi$) at T = 45 K for the OP samples and 8 K ( and 45
K) for the OD samples. We stress that very high-resolution and
high counting statistics are required for the analysis which
follows below. To obtain the images in Fig. 1 and 2, a typical
measurement time of  $15-20$ hours is needed. Each cut is taken in a
series of 20-min-long scans to ensure that no significant changes
occur in the spectra. Sample aging can also be checked by
comparing the peak heights of the energy distribution curve (EDC)
before and after the measurement. We discarded the scans when the
peak height changed by more than 5\%. Fig. 1 shows an ARPES
spectrum of the OD sample along the nodal direction. To isolate
the structure of electron-boson coupling, we extract the peak
position and line width from momentum distribution curves (MDC) by
fitting to Lorentzian curves. We note that the difference of the
peak-position plots at 8 K and 45 K is insignificant (see the
inset in Fig. 1c). Therefore, we will use the less noisy 8K data
for the self-energy analysis. In Fig. 1c, the high quality
measured electron dispersion shows clearly a kink around $70-75$
meV. The corresponding change in the MDC width is consistent with
the energy position of the kink.

To obtain information on the electron-boson coupling, our analysis
is aimed at isolating the strength and shape of the real part of
electron self energy Re($\Sigma$). To extract Re($\Sigma$), we
subtract a bare dispersion from the measured one; the bare
dispersion for OD sample is shown in Fig. 1c. The bare dispersion
is approximated with a second-order polynomial where the fitting
parameters are chosen such that the bare and experimental
dispersion are in agreement at high binding energies, resting on
the assumption that the bosonic couplings at high energies are
diminishing such that the bare and renormalized bands merge. We
assume this to be the case in the $150-250$ meV range. Notice that
in this way the large but featureless self-energies associated
with the electron-electron interactions are absorbed in the bare
dispersions.

\begin{figure}[t]
\includegraphics [width=3.0in, clip]{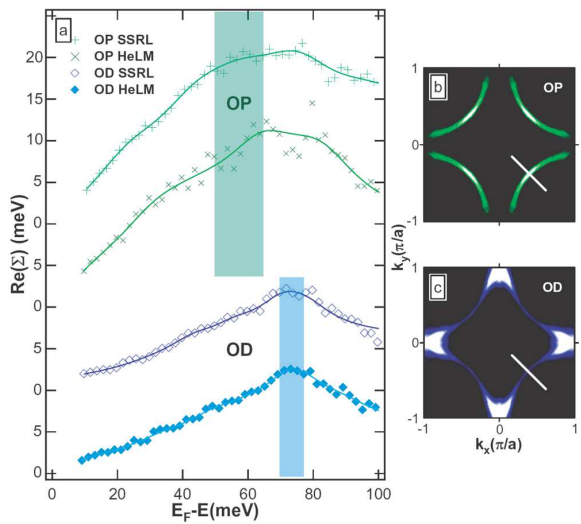}
\caption{\label{fig2:dopingdependent} (a) Comparison of
Re($\Sigma$) for OD and OP samples, showing doping dependence in
overall coupling strength and relative strength in mode energies.
Note that SSRL and HeLM mean, measured with the 23.7eV synchrotron
light and the He I light and the blue and red areas show the
dominated feature in OD and additional feature in OP,
respectively. (b) and (c) are respectively symmetrized schematic
Fermi surface maps of OD and OP samples.}
\end{figure}

By applying this procedure to the ARPES electron dispersions of
the OD and OP samples, we arrive at the main result of this paper:
the difference in the extracted Re($\Sigma$) at the two dopings
(Fig. 2a). To confirm this result, we have performed two sets of
experiments in two different ARPES systems under different
experimental conditions, especially so with regard to the photon
polarization. As shown in Fig. 2a, the two data sets show a  good
overall agreement, adding confidence to this result.

Re($\Sigma$), as extracted via this procedure, is seen to change
drastically from the OP to the OD regime (Fig. 2a). First, its
overall magnitude is significantly reduced, in accordance with
earlier observations in LSCO \cite{MultiMode:XJZhou} and
Bi$_{2}$Sr$_{2}$CaCu$_{2}$O$_{8}$ (Bi2212)
\cite{Mode:Johnson,Mode:Lanzara}. Taking the area underneath the
curve as an indication of the coupling strength we find a change
from 730 (OP) to 340 (OD) (meV)$^2$.

The surprise is that Fig. 2a reveals a qualitative change occuring
in the energy dependence of Re($\Sigma$) as a function of doping.
One obtains the impression that the OD self-energy is dominated by
a feature centered at $70-75$ meV. In the OP self-energy there
appears to be much more weight in the $30-60$ meV range, which has
largely decreased in the OD system.  As noted by Zhou \emph{et
al.} \cite{MultiMode:XJZhou} the self-energy of the OP sample is
reminiscent of a spectrum of modes,  and we take the large doping
induced changes as support for this claim.

\begin{figure}[t]
\includegraphics [width=3.0in, clip]{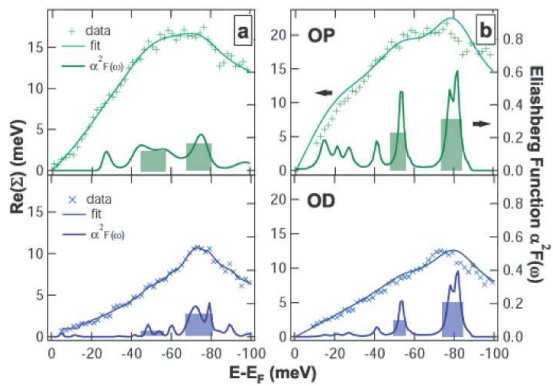}
\caption{\label{fig3:lossfunction} (a) Eliashberg function and
self-energy for OP (top row) and OD (bottom row) samples
calculated by MEM analysis. (b) Comparison of measured
Re($\Sigma$) with a simulation using the original (for OP) and
'screened' (for OD) loss-function spectrum. Note that shaded areas
show the similar sizes in area of the main features in the two
analysis.}
\end{figure}

To obtain an impression of the form of the Eliashberg function
\cite{eliashberg:Grimvall}, $\alpha^2F(\omega)$, corresponding to
the extracted self-energy, we employ the inversion based on the
maximum entropy method (MEM) described in ref.\cite{MEM:Junren}.
In Fig. 3a we show the result of the MEM-analysis. We checked the
self-consistency of this result by comparing the width $\Gamma$ (
Im($\Sigma$)= $(\Gamma/2)v_\circ$) as computed from the
MEM-$\alpha^2F(\omega)$ with the width derived from the
experimental MDC.  In accordance with the expectations, the
MEM-$\alpha^2F(\omega)$ is characterized by a high energy
structure around $\omega = 70-80$ meV. However, in the OP system
one finds in addition  much weight in $\alpha^2F(\omega)$ in the
$\omega = 30-60$ meV range which has largely decreased in the OD
system.

The rather detailed information we have obtained as a function of
(over)doping is quite informative regarding the physical nature of
the modes coupling to the electrons. The structured nature of the
self-energy is strongly suggestive of phonons because phonons
could provide a spectrum of modes at all dopings, and in this
regard they are unique. The question immediately arises, how can
one explain the gross changes we observe as a function of doping
in terms of phonons? It implies a drastic reorganization of the
way electrons couple to phonons.

In the range of frequencies of present interest  ($30-100$ meV)
one is dealing predominantly with motions of the light oxygen
ions.  A first candidate is the breathing phonons in the range of
$70-90$ meV, involving planar oxygen motions in ab plane. These
show doping induced anomalies and it was argued that these have to
contribute to the self-energy in nodal directions
\cite{Mode:Lanzara,B1gMode:Tanja,B1gMode:Tom}. However, these
anomalies are not known to change significantly in this doping
range \cite{HalfBreathing:Fukuda} and these planar phonons are
therefore less likely to be responsible for the dramatic change in
the self-energy in this doping range.  We will now present
evidence that this gross change can be interpreted as an effect of
metallization occurring along the c-axis going from OP to OD.

The c-axis electron energy loss function Im$(-1/\epsilon_c
(\omega))$ determined by Tsvetkov {\em et al.}
\cite{LossFunction:Marel} from the c-axis optical response  of a
superconducting Bi2201 sample turns out to be a remarkably
accurate model for the Eliashberg function ($\alpha^2F(\omega)$)
needed for the self-energy of our OP system (see Fig. 3b). The
c-axis loss function reflects the various electromagnetically
active ionic motions along the c-direction at zero momentum and
these are assigned as follows \cite{LossFunction:Marel}: the low
energy structures $< 30$ meV can be ascribed to the motions of the
heavy atoms, the peaks at 40 and 50 meV are assigned to out of
plane motions of planar oxygens, and the high energy peak at 80
meV is mostly due to out of plane motions of the apical oxygens
which could have a character of apical and breathing phonons, of
in-plane oxygen, but with a satellite due to structural
distortion.

Furthermore, we can also fit the self energy of the OD system by
multiplying this same model for $\alpha^2F(\omega)$ with a `filter
function', $\omega^2 / ( \omega^2_{scr,c} + \omega^2)$, which
reduces the spectral weight below $\omega_{scr,c}$. The
implication from the addition of this filter function to the
'unscreened' $\alpha^2F(\omega)$ is that phonons below
$\omega_{scr,c}$ are screened out when the c-axis metallicity sets
in. The reasonable fit shown in Fig. 3b is obtained using
$\omega_{scr,c} = 60$ meV as the single free parameter, besides
the overall scale parameter $g \approx 0.7$ for both OP and OD
where $\alpha^2F(\omega) = g$ Im$(-1/\epsilon_c(\omega))$.

How can it be that the empirical correlation between the
\emph{c-axis} loss function and the electron self-energy along the
nodal direction, assumed to be due to electron-phonon (EP)
coupling, works so well for OP system? The EP coupling in oxidic
insulators is gigantic by the standards of metals
\cite{MullerBednorz}, as it resides in the long range
electrostatic interactions between the electron charge and the
highly polarizable lattice.   The electron self-energy due to this
non-screened, and therefore polar, EP coupling can be approximated
by the electron energy loss function -- a detailed theoretical
discussion of this point will be presented in a following paper
\cite{nonth}.   The connection to the {\em c-axis} loss function
is then justified by noting that the polar EP interaction is
screened for any sizable planar momentum transfer. When we view
cuprates as a stack of metallic sheets, it has been shown that for
phonons with 3D momentum $\vec{q} = \vec{q}_{ab} + \vec{q}_c$, a
characteristic frequency $\Omega_{sc}$ can be identified for the
three dimensional problem \cite{PlasmonInOut:Marel, Falikov},
$\Omega^2_{sc} ( \vec{q} ) = (q_c^2/q^2) \omega_{scr,c}^2 +
(q_{ab}^2/q^2) \omega_{p,ab}^2$ such that all phonons with
frequency $\omega_{ph} (\vec{q}) < \Omega_{sc} (\vec{q})$ can be
regarded as screened. Since the planar plasma frequency
$\omega_{p,ab}$ is large compared to the c-axis characteristic
screening frequency $\omega_{scr,c}$, this implies that only
phonons with $q_{ab} \simeq 0$ can contribute to the polar
couplings.  It is important to note that the electron energy loss
function, as determined by optics, only measures the $q
\rightarrow 0$ part of the loss function.  And, the most important
modes do not disperse much as a function of c-axis momentum, so
the optical loss function should be quite representative for this
kinematic regime. However, we note that we do not claim that the
coupling to planar phonons is non-existent; these modes may well
lie hidden in the background and be less likely to be the cause
for the strong change in this doping range.

The next question is why the self-energy changes so much in this
doping range? Although we are not aware of systematic measurements
of the doping dependence of the electrodynamical properties along
the c-axis of Bi2201, comparison with other cuprates (the LSCO
\cite{LSCO:Tajima} and YBCO \cite{YBCO:Tajima} systems) suggests
that the \emph{screened} loss function for OD system is
reasonable. Optical measurements have revealed that the
metallization of the c-axis is primarily driven by a drastic
decrease in the c-axis charge relaxation rate $\Gamma_c$ changing
from strongly overdamped to moderately overdamped in going from OP
to OD \cite{PlasmonInOut:Marel}. Further, such a change can
account in detail for the changes in the self-energy, assuming
that the Bi2201 system behaves similarly in this regard to the
LSCO system in the doping range $x=0.15 - 0.30$ \cite{nonth}.
Bi2201 does not seem to be an exception, given that for instance
its c-axis resistivity reduces upon doping \cite{CaxisRes:Ono} and
shows a metallic behavior in the OD sample \cite{CaxisRes:Chong}.

In conclusion, we have found rather dramatic changes in the
self-energy of nodal electrons between OP and OD samples,
reflecting a change in the coupling of electron and bosonic modes.
This change is manifested in the clear disappearance of coupling
of the modes in the intermediate energy range ($30-60$ meV). We
interpret this effect as caused by a dramatic change in the
coupling to c-axis phonons, turning from polar- into metallic in
this doping regime. We stress that the presence of these polar EP
interactions in an otherwise metallic system is highly unusual and
there have to be more surprises in store.

We thank D. van der Marel and A. Damascelli for enlightening
discussions and X.J. Zhou and W.L. Yang for helping with early
experiments. SSRL is operated by the DOE Office of Basic Energy
Science under Contract No. DE-AC03-765F00515. ARPES measurements
at Stanford were supported by NSF DMR-0304981 and ONR
N00014-98-1-0195. W.M. acknowledges DPST scholarship for the
support. T.P.D. would like to thank ONR N00014-05-1-0127, NSERC,
and Alexander von Humboldt foundation. J.Z. acknowledges the
support by the Fulbright foundation in the form of a senior
fellowship.


\begin{thebibliography}{99}

\bibitem{Group:Review}
A. Damascelli, Z. Hussain, and Z.-X. Shen, Rev. Mod. Phys.
\textbf{75}, 473 (2003).
\bibitem{Mode:Pasha}
P.V. Bogdanov \emph{et al.}, Phys. Rev. Lett. \textbf{85}, 2581
(2000).
\bibitem{Mode:Kaminski}
A. Kaminski \emph{et al.}, Phys. Rev. Lett. \textbf{86}, 1070
(2001).
\bibitem{Mode:Johnson}
P.D. Johnson \emph{et al.}, Phys. Rev. Lett. \textbf{87}, 177007
(2001).
\bibitem{Mode:Lanzara}
A. Lanzara \emph{et al.}, Nature \textbf{412}, 510 (2001).
\bibitem{mode:Fink}
T.K. Kim \emph{et al.}, Phys. Rev. Lett. \textbf{91}, 167002
(2003).
\bibitem{Mode:Gromko}
A.D. Gromko \emph{et al.}, Phys. Rev. B \textbf{68}, 174520
(2003).

\bibitem{Mode:Gweon}
G.H. Gweon \emph{et al.}, Nature \textbf{430}, 187 (2004).

\bibitem{Norman:SpinAndTightbinding}
M. Eschrig and M.R. Norman, Phys. Rev. B \textbf{67}, 144503
(2003).

\bibitem{B1gMode:Tanja}
T. Cuk \emph{et al.}, Phys. Rev. Lett. \textbf{93}, 117003 (2004).

\bibitem{Bi2223:Weisheng}
W.S. Lee \emph{et al.},  to be published.

\bibitem{MultiMode:XJZhou}
X.J. Zhou \emph{et al.}, Phys. Rev. Lett. \textbf{95}, 117001
(2005).

\bibitem{Polaron:Kyle}
K. M. Shen \emph{et al.}, Phys. Rev. Lett. \textbf{93}, 267002
(2004).

\bibitem{MagneticResonance:Keimer}
H. He \emph{et al.} Phys. Rev. Lett. \textbf{86}, 1610 (2001). B.
Keimer \emph{et al.} Physica C, \textbf{341}, 2113 (2000).

\bibitem{PlasmonInOut:Marel}
D. van der Marel and J.H Kim, J. Phys. Chem. Sol. \textbf{56},
1825 (1995)

\bibitem{smallq:Tom}
T.P. Devereaux \emph{et al.}, to be published.

\bibitem{eliashberg:Grimvall}
G. Grimvall and E. Wohlfarth, The Electron-Phonon Interaction in
Metals, (North-Holland, New York, 1981).

\bibitem{MEM:Junren}
J.R Shi \emph{et al.}, Phys. Rev. Lett. \textbf{92}, 186401
(2004).

\bibitem{B1gMode:Tom}
T.P. Devereaux\emph{et al.}, Phys.Rev.Lett.\textbf{93},117004
(2004).

\bibitem{HalfBreathing:Fukuda}
T. Fukuda \emph{et al.}, Phys. Rev. B \textbf{71}, 060501 (2005).

\bibitem{LossFunction:Marel}
A.A. Tsvetkov \emph{et al.}, Phys. Rev. B \textbf{60}, 13196
(1999).

\bibitem{MullerBednorz}
K.A. Muller and J.G. Bednorz, Science \textbf{237}, 1133 (1987).

\bibitem{nonth} W. Meevasana \emph{et al.}, to be published.

\bibitem{Falikov} H. Morawitz {\em et al.}, Z. Phys. B {\bf 90},
277 (1993) and references therein.

\bibitem{LSCO:Tajima}
S. Uchida \emph{et al.}, Phys. Rev. B \textbf{53}, 14558 (1996).

\bibitem{YBCO:Tajima}
S. Tajima \emph{et al.}, Phys. Rev. B \textbf{55}, 6051 (1996).

\bibitem{CaxisRes:Ono}
S. Ono and Y. Ando, Phys. Rev. B \textbf{67}, 104512 (2003).

\bibitem{CaxisRes:Chong}
I. Chong \emph{et al.}, Physica C \textbf{290}, 57 (1997).

\end{thebibliography}
\end{document}